# A 4D synchrotron X-ray tomography study of the formation of hydrocarbon migration pathways in heated organic-rich shale

*Hamed Panahi[1,2], Maya Kobchenko[1], François Renard[1,3], Adriano Mazzini[1], Julien Scheibert[1,5], Dag K. Dysthe[1], Bjorn Jamtveit[1], Anders Malthe-Sorenssen[1] and Paul Meakin[1,4,6]*
1. Physics of Geological Processes, University of Oslo, Norway
2. Statoil, Oslo, Norway
3. ISTerre, University of Grenoble I & CNRS, BP 53, 38041 Grenoble, France
4. Idaho National Laboratory, Idaho Falls, USA
5. Ecole Centrale de Lyon, Laboratoire de Tribologie et Dynamique des Systèmes Lyon, France
6. Institute for Energy Technology, Kjeller, Norway

**Abstract**
Recovery of oil from oil shales and the natural primary migration of hydrocarbons are closely related processes that have received renewed interests in recent years because of the ever tightening supply of conventional hydrocarbons and the growing production of hydrocarbons from low permeability tight rocks. Quantitative models for conversion of kerogen into oil and gas and the timing of hydrocarbon generation have been well documented. However, lack of consensus about the kinetics of hydrocarbon formation in source rocks, expulsion timing and how the resulting hydrocarbons escape from or are retained in the source rocks motivates further investigation. In particular, many mechanisms for the transport of hydrocarbons from the source rocks in which they are generated into adjacent rocks with higher permeabilities and smaller capillary entry pressures have been proposed, and a better understanding of this complex process (primary migration) is needed. To characterize these processes it is imperative to use the latest technological advances. In this study, it is shown how insights into hydrocarbon migration in source rocks can be obtained by using sequential high resolution synchrotron X-ray tomography. Three-dimensional (3D) images of several immature "shale" samples were constructed at resolutions close to 5 micrometers. This is sufficient to resolve the source rock structure down to the grain level, but very fine grained silt particles, clay particles and colloids cannot be resolved. Samples used in this investigation came from the R-8 unit in the upper part of the Green River Shale, which is organic rich, varved, lacustrine marl formed in Eocene Lake Uinta, United States of America. One Green River Shale sample was heated in-situ up to 400°C as X-ray tomography images were recorded. The other samples were scanned before and after heating at 400°C. During the heating phase, the organic matter was decomposed, and gas was released. Gas expulsion from the low permeability shales was coupled with formation of microcracks. The main technical difficulty was numerical extraction of microcracks that have apertures in the 5 to 30 micrometer range (with 5 micrometers being the resolution limit) from a large 3D volume of X-ray attenuation data. The main goal of the work presented here is to develop a methodology to process these 3D data and image the cracks. This methodology is based on several levels of spatial filtering and automatic recognition of connected domains. Supportive petrographic and thermogravimetric data were an important complement to this study. An investigation of the strain field using two-dimensional image correlation analyses was also performed. As one application of the four-dimensional (4D, space + time) microtomography and the developed workflow, we show that fluid generation was accompanied by crack formation. Under different conditions, in the subsurface, this might provide paths for primary migration.

**Key words** 4D microtomography, 3D image processing, shale, strain field analysis, kerogen, petroleum generation, primary migration, petrography, thermogravimetry

# 1. Introduction
Over the past few years X-ray microtomography or X-ray computed tomography (CT) - a non-destructive technique - has become increasingly more important and widely applied to the study of rock morphology, microtexture, transport properties and fracturing (Lindquist et al., 2000; Arns et al., 2001; Mees et al., 2003). The internal structure of the rock samples was determined from variations in the X-ray attenuation, which depend on the atomic structure and density. One-dimensional (1D) or two-dimensional (2D) radiographs are acquired by stepwise rotation of either the X-ray source/detector or the sample followed by mathematical reconstruction of the horizontal cross-sections, which are perpendicular to the axis of rotation. It is now common knowledge that CT registers the attenuation of the X-rays which is material and X-ray energy dependent. The inelastic scattering of X-ray photons in matter results in a decrease in energy (an increase in wavelength), and photoelectric attenuation is also important, especially at low energies. Unless the X-rays are produced by radioactive decay, they are always polychromatic with a wide range of energies (wavelengths). This complicates quantitative analysis and creates artifacts in the CT images, due to the stronger attenuation of X-rays with lower energies. Monochromatization by diffraction eliminates these problems, but it involves a great decrease in intensity and is therefore only feasible for high intensity X-ray sources, such as linear electron accelerators and large synchrotron "light" sources (Mees et al., 2003). Because of recent advances in computer processing power, high accuracy X-ray CT is becoming more commonplace in rock physics laboratories around the world.

Most studies have focused on the analysis of static 3D tomograms. However, with increased resolution and decreased scan duration, made possible by high intensity, highly collimated, quasi-monochromatic X-ray beams, it is now feasible to investigate time-dependent processes. We have applied this methodology to study the coupling between fluid generation and fracturing in Green River Shale – an experimental model for primary migration.

A number of mechanisms have been proposed to explain primary migration, the generation of hydrocarbon fluids in source rocks and the expulsion of these fluids into nearby secondary migration pathways. These mechanisms include migration through a continuous oil saturated kerogen network (McAuliffe, 1979), migration of microdroplets, migration of micelles (Cordell, 1973), migration of oil emulsion, migration of oil dissolved in water (Bray and Foster,1980) and migration of oil dissolved in gas (Price,1989). However, there is still no consensus concerning the controlling mechanisms (Hermanrud, 1991). In all likelihood, several mechanisms play a role in primary migration, and their relative importance depends on the initial organic and inorganic composition of the shale, the heterogeneity, the stage of maturation, the temperature/pressure history and other factors.

The idea that microcracks play an important role in primary migration is supported by observations that cracks at all scales pervade mature source rocks (Chanchani et al.,1996; Lash and Engelder, 2005) and laboratory specimens after experiments in which shales with a significant organic content were thermally decomposed under confinement (Tissot and Pelet,1971). In at least some source rocks, microcracks are expected to heal rapidly after they have formed (Mann, 1994), and characterization of the crack network in mature source rocks may lead to substantial underestimation of the role played by microcracks. The mechanical properties of the source rock may be altered during the maturation process, and the roles played by fracturing and crack healing depend on these properties and the rate of fluid production as well as the nature of the fluid.

Flow in cracks or other high permeability and low capillary pressure pathways appear to be essential to support the migration of hydrocarbon over large distances from the interior of thick source rock beds to secondary migration pathways. Except possibly for methane, the diffusion coefficients are too small, and the solubilities in water are too low (England *et al.*, 1987; Hunt,1996) for diffusion through the organic and/or aqueous phases to be sufficient by itself, and the permeabilities of, typically very fine grained, source rocks are too low and the capillary pressures are too high for fluid flow to be sufficient without fracturing. This has led to the idea that microcracks act as fluid pathways during primary migration (Snarsky, 1961, Tissot and Welte, 1978). Nevertheless, most experiments performed to obtain a better understanding of primary migration (Lafargue *et al.*, 1993; Rudkiewicz *et al.*, 1993) do not address the role played by cracks. Of course, other mechanisms are required to transport the hydrocarbon fluids from where they are formed to the microcracks, but mechanisms such as diffusion and viscous flow through very low permeability rock matrices may be more effective over these very short distances.

To be able to properly characterize microscopic and macroscopic transport phenomena in tight rocks including shales, it is imperative to obtain quantitative information about the evolution of microcracks. Much progress has been made in digital data processing and imaging techniques and this has made it possible to study processes at much finer scales. Several methods can be used to visualize the pore structure in two dimensions (planar cuts through three-dimensional objects). SEM (Scanning Electron Microscopy) (Danilatos, 1988), TEM (Transmission Electron Microscopy) and XRF (X-ray Fluorescence Imaging) have been widely used to obtain 2D images with micron and sub-micron resolutions. Constructing a 3D representation of a sample from multiple 2D micrographs is challenging and cannot reproduce all the sample heterogeneities. If the sample must be removed to prepare a new surface or thin section, it can be very difficult to reposition the sample and obtain a new image with the same orientation as previous images and a position that is known relative to the positions of the previous images with a high enough accuracy to prevent distortion of the final 3D image. Moreover, the process used to obtain thin sections or prepare flat surfaces for TEM, XRF, and SEM scans requires destruction of the sample. For these reasons, we used X-ray micro-computed tomography as a non-destructive 3D imaging technique. Recently, focused ion beam scanning electron microscopy (FIB-SEM) has been used to obtain 3D images of shales and mudstones (Schiffbauer and Xiao, 2009). In FIB-SEM, the focused ion beam is used to perform nanotomography by alternately milling or cross-sectioning the sample and recording an SEM image. The SEM images can be used to construct a very small scale 3D image, which complements the information obtained from coarser scale 3D X-ray tomography. However, this destructive technique cannot be used to directly determine how the structure changes in time under constant conditions or as conditions are changed.

X-ray microtomography, which was introduced earlier, has been used for a variety of geological applications including measuring the microstructures and mineral contents of rocks (Weavers et al., 2000; Lindquist et al. 2000), determining the spatial characteristics and distributions of minerals and investigating processes such as fracturing, dissolution and precipitation (Renard et al., 2004, 2009). Determination of pore network geometry to calculate permeability and porosity is widely used despite the need for high speed large memory computing systems (Arns et al., 2001; Renard et al., 2004).

In the present study, we imaged the nucleation and growth of micro cracks in tight rocks; including mudstones and shales on scales down to the experimental resolution limit. A large number of samples from different geographical locations were imaged using computed X-ray microtomography in order to investigate geometrically complex cracks and the growth of cracks inside optically opaque materials. The resulting 3D volumes were represented by large 3D voxel maps, which required fairly intense data processing. A customized workflow was developed for processing the tight rock tomograms, with the objective of reducing the processing time. To complement this study, some samples were further investigated using SEM and thermogravimetric analysis. This helped us to identify correlations between the volumes in which organic material is present and the volumes in which microcrack nucleation and growth (underlying strain field development) occur, and determine the temperatures over which kerogen transformation happened.

## 2. Materials and Methods

### 2.1 Shale samples

Shale samples were collected from the organic rich R-8 unit in the upper part of the Green River Shale Formation, USA; (see **Table 1**). The Green River Shale is a marl, which was deposited in Eocene Lake Uinta. The organic matter has undergone conversion to type I kerogen, but there has been little or no thermal maturation of the kerogen. The R-8 unit, from which the Green River Shale samples were obtained is of potential commercial interest because if its high oil yield (GRSII and GRSIV samples in **Table 1**).

## 2.2 Shale geochemical and petrophysical characterizations

Several samples were chosen for geochemical and petrophysical characterizations before acquiring X-ray tomographic images (tomograms). Total carbon content and organic and inorganic carbon ratio measurements, optical observations on thin sections and chemical analysis using an electron microprobe were performed to characterize the mineralogical composition, the organic content and the microstructure of the shales.

### 2.2.1. Microstructure study

Several thin sections, prepared from the Green River Shale sample adjacent to the 4mm × 4mm cores extracted from the larger shale samples for X-ray tomography, were used for scanning electron microscopy (SEM), including back scattered electron imaging (BSE) and energy dispersive X-ray spectroscopy (EDS) imaging. Analysis of the data acquired in these experiments allowed us to identify the elements and the types of minerals present, and also map the organic and inorganic (carbonate) carbon. **Figs. 1a and 1b** show BSE images of two Green River Shale samples in which organic carbon and pyrite ($FeS_2$) are clearly visible owing to their grey level contrast.

### 2.2.2 Organic content

The presence of organic carbon was further investigated using total organic carbon (TOC) and total inorganic carbon (TIC) analyses (**Table 1**), which showed, that these samples contained significant amounts of organic carbon. In addition, aerobic and anaerobic thermogravimetric analyses were performed on 20 mg powder samples to investigate the presence of volatile fluids such as water, organic substances and inorganic minerals that undergo thermal decomposition to form volatile products such as carbon dioxide. We used thermogravimetric analysis with simultaneous differential thermal analysis (ATG/SDTA) with a Mettler Toledo ATCG/SDTA851 thermogravimetric analyzer. This technique continuously measures the weight of a sample during heating from 20°C to 1200°C. The release of volatile components during heating occurs in stages, and the temperatures associated with each stage provide an indication on the nature of the component(s) that evaporates and/or decomposes. Thermogravimetry was also coupled with gas chromatography/mass spectrometry (GC/MS) using an Agilent 5890 GC/5973 mass spectrometer system to analyse the gases (water, $CO_2$, and organic volatiles) that were driven off. The thermogravimetric analyses were performed either under air, or in an oxygen-free nitrogen environment at atmospheric pressure.

For immature samples of Green River Shale, several steps of weight loss could be identified. There was an interval between 250°C and 380 °C which corresponds to dehydration of clay minerals and another distinct step in the 380°-470°C temperature range, which corresponds to the release of various organic molecules. This was observed for the Green River Shale samples (**Fig. 2**). The X-ray tomography analysis, discussed in the following section, indicates that this loss of weight is accompanied by the formation of microcracks in the sample. The weight loss occurred in the 750°C to 850°C range corresponds to the decomposition of carbonate minerals.

## 2.3 X-ray tomography acquisition and 3D reconstruction

Cylindrical core samples (4mm x4 mm) were extracted from larger shale samples, perpendicular to the stratification, and were imaged in 3D using X-ray tomography. Two X-ray computed micro-tomography campaigns were conducted at the X02DA TOMCAT beamline at the Swiss Light Source (SLS) in Zürich, Switzerland, and at the ID19 beamline at the European Synchrotron Radiation Facility (ESRF) in Grenoble, France. The voxel resolutions of the images were 5.05 micrometer at SLS and 4.91 micrometer at ESRF. In both cases, collimated 20 keV X-ray beams were used. Unlike the ring artifacts in the images obtained from the ESRF samples, most of the samples imaged at SLS had ring artifacts with a wide range of grey scale values, which made their removal without removing some of the sought after features, such as micro cracks, impossible.

In addition, one of the Green River Shale samples was imaged while its temperature was continuously increased from 60ºC to 400 ºC. This was performed using an in-situ oven available on site at ESRF (Terzi et al., 2009). The oven was controlled by a computer and a heat ramp of 1°C/minute was programmed between 60°C and 400°C, while 3D scans were continuously acquired. On average, each scan required approximately 20 minutes, and 68 images were acquired during the heating phase explaining that temperature remained constant at 400°C after acquiring the 17$^{th}$ image for the additional 51 scans. Therefore, it was possible to follow the evolution of this sample during the decomposition of organic matter.

Most of the analyses were carried out on this sample, because it was possible to track crack nucleation and growth as resolved at 5 µm and characterize the cracks at many steps during their quasi-static propagation. Other 3D X-ray attenuation images were analyzed in detail to determine the orientation, density, spacing and geometries of the cracks and study their displacement/strain fields.

Before the images can be utilized for further processing, and eventually scientific interpretation, there is one intermediate step which was done at the synchrotron facilities in Switzerland and in France. Full 3D representations of the samples were constructed from 2D X-ray images using the projection slice theorem (Bracewell, 1990; Levoy, 1992) with enhanced features for more accurate reconstruction (Deans,1983). Several artifacts, which have been discussed in the literature, including beam hardening, hot points and ring artifacts were reduced by filtering the raw data. However, ring artifacts were apparent in some of the reconstructed 3D data. The ring artifacts from analysis of the ESRF data could be filtered out in the later stages of image processing by removing the high contrast voxels, because their grey scale values were at one end of the spectrum in multiple slices of the imaged samples. This was not possible for the SLS rings artifacts because they had a wider range of grey-scale values, which spread well into the range of grey-scale values associated with the features of interest.

**3. Image Analysis**

The two-dimensional and three-dimensional image analyses of the shale samples were utilized for different purposes. The two-dimensional images were used for 2D image correlation analysis, and this allows the deformation field to be determined without image segmentation. The three-dimensional images were used to study fracture nucleation, growth and coalescence, and the evolving visible patterns provided information that was helpful in explaining the phenomena behind the fracturing behavior.

**3.1 Analysis of the strain field using 2D image correlation**

Digital Image Correlation (DIC) (Hild and Roux, 2006) analysis was performed on two orthogonal vertical slices of the tomographic images obtained as a Green River Shale sample was heated. In each of these two planes, the evolution of both the vertical and horizontal strain was determined as a function of temperature in the following way.

Each image (deformed image) was compared with the immediately preceding image. Around each pixel of the deformed image, a sub-image of 25 times 25 pixels was extracted and correlated to the corresponding sub-image in the reference image (the immediately preceding image). A 25 by 25 correlation matrix was obtained by multiplying the Fourier transforms (obtained using a Fast Fourier Transform algorithm) of both subimages (one being rotated by 180 degrees in order to account for complex conjugation) before taking the inverse Fourier transform of the result. The maximum element of the correlation matrix was first located. A sub-pixel determination of the location of the maximum was then obtained by finding the maximum of the best (least mean square deviation) parabola representing the 3 by 3 correlation sub-matrix centered on the element with the maximum correlation. The $x$ and $y$ distances between this maximum and the center of the correlation matrix were assumed to be the $x$ and $y$ components of the displacement vector, for this pixel, from the reference to the deformed image.

The size of the sub-images was chosen to be 25 after several sizes were tested. Prior to crack nucleation, displacements up to 5-6 pixels were measured between two successive images, and since the maximum displacement measurable with a sub-image of size N is N/4, the minimum possible sub-image size is 6x4=24 pixels. We then tested three different sizes: 25, 37 and 49 pixels, which all gave results very similar to that shown in **Fig. 3**, within variations less than 11% for all temperatures below 350 $^O$C. This result strongly suggests that the correlation analysis is almost independent of the size of the sub-image, provided that it is ≥25. We then selected the smallest size among these three, so that the spatial resolution of the displacement field was optimal, with minimal computational effort.

The correlation analysis provides the evolution of the incremental displacement field as the sample deforms due to heating (**Fig. 3**). Limitations in the maximum displacement measured (see discussion) prevented us from correlating all successive images to the first image. In other words, we only had access to displacement increments between successive image pairs, not to the absolute displacements. The total displacement can, of course, be obtained as the sum of all previous increments, but with the drawback that the noise level of the results will increase in proportion to the square root of the number of added incremental displacement fields.

For each image, we determined that the *x* displacement averaged along the *y* direction as well as the *y* displacement averaged along the *x* direction are linear function of *x* and *y*, respectively (**Fig. 4b**). This indicates that, upon heating, the thermal expansion strain increase is essentially homogeneous in both directions for this unconfined sample. The strain increment can therefore be represented by two scalar values, one for each direction, which are plotted as a function of temperature in **Fig. 4a**. The strain is anisotropic, and the horizontal strain increment (along the stratification) is significantly smaller (by a factor of about 2 to about 5 depending on the temperature, see **Fig. 4a**) than the vertical one (perpendicular to the stratification). Their values vary with temperature, indicating that the linear thermal expansion coefficient (strain increment per unit degree) is itself a function of temperature. It is roughly constant from 60 °C to around 290 °C, with typical values of about 5e-5 /degree in the vertical direction and about 2e-5 /degree in the horizontal direction Between 290 and 350 °C, the vertical linear thermal expansion coefficient increases quite rapidly with increasing temperature. The total strain at about 350 °C, obtained by adding strain increments (**Fig. 4b**) reaches ≈2% in the vertical direction and ≈0.6% in the horizontal direction. All these results were almost the same for the two slices.

Above 350°C, cracks form and grow. These new features in the images have no equivalent in the previous images and this prevented us from using the correlation analysis for T>350°C.

## 3.2 3D Image analysis

A methodology was developed for 3D image analysis including several steps which should be carefully followed to obtain the best possible quantification of the crack network. Despite the application of powerful computer processing units, the size of the samples made the processing very slow. The selection of the best methods at each step to remove most of the unwanted data in the form of noise or features which are not of interest while keeping small features of interest was challenging.

### 3.2.1 Image processing

The raw 3D X-ray tomography volumes must be analyzed to provide images, extract features of interest, and perform qualitative or quantitative analyses that can be used for interpretation. This analysis can be time-consuming (can take up to one day) given the large size of each data set (each volume is represented by either 2048x2048x256 or 830x830x830 voxels, coded as 8 or 16 bit grey levels) and the need to develop a workflow for all the steps necessary for image processing, from filtering to 3D rendering and quantification of deformation patterns. However, this kind of processing is becoming more and more mainstream. Moreover, characterization of the nucleation, growth and healing of cracks during thermal alteration was an important aspect of this study, and to do this, the microcracks must be distinguished from other low X-ray attenuation objects such as kerogen lenses and imaging artifacts. A filtering scheme, performed in several steps, that allows segmentation of the cracks is presented here **(Fig. 5)**. This is applicable to images of very tight rocks with distinguishable small scale features (in this case micro cracks). The software package Avizo Fire was used to develop the image processing scheme. As shown in **Fig. 6**, small crack apertures (up to 5 voxels/25 micrometers) and the wide range of crack sizes make segmentation challenging. Sample displacement is also an issue that should be addressed during image reconstruction or at a later stage during data processing. For our images, displacement was not a significant issue because the samples that were analyzed did not undergo displacements that were large enough to affect the accuracy of the method adopted.

### 3.2.2 Primary filtering, pre-processing and masking

The microstructure of the rock is readily apparent in the primary 3D images. Apart from the typical imaging artifacts already mentioned, there can be other issues like uneven illumination of the samples during the imaging process, Gaussian noise, "salt-and-pepper" noise, quantization noise, shot noise and sub-voxel features below the imaging resolution. While there are standardized techniques to deal with some of these sources of image noise, a number of them require user interaction and additional information about the object(s) of interest in order to properly distinguish between the noise and the signal.

A variety of filtering algorithms were tested and their outputs were compared to determine which gave the best results.

These included: 1. three noise reduction filters (minimum, median and maximum) which replace the value of a pixel by the smallest, median and largest value of the neighboring pixels using an NxN mask; 2. a Sobel edge detection filter; 3. a re-sampling/low pass filter based on convolution with a Lanczos kernel; and 4. an edge-preserving Gaussian filter.

The Sobel filter is a rotation invariant edge detection filter. This filter convolutes the image with 4 different kernels representing horizontal, vertical and two arbitrary diagonal orientations, and each kernel consists of a combination of Gaussian smoothing/filter and differentiation in the proper orientation, The Lanczos filter indicates which features in the original data, and in what proportion make up each feature in the final data, a statistical feature detection filter. This filter calculates the n-th centralized moment of the data in a gliding window.The edge-preserving Gaussian filter gave the most satisfactory results. It preserved the outlines and areas of features of interest while removing most of the artificial features and/or natural features that were not of interest. A special type of Gaussian filter was used to preserve edges while smoothing the overall 3D image. This filter has an effect similar to the physical process of diffusion, which converts a delta function into a Gaussian (it smoothens the grey level intensity differences between neighboring voxels). However, it does not smear out edges because the "diffusion coefficient" is reduced, to zero, as edges are approached. The presence of distinctive particles like pyrite particles helped us to select, customize and calibrate the filtering technique and the relevant parameters. The blank part of the image as well as parts of the samples which were chipped off during the coring process, and the central part of the sample with some noise, which established artificial vertical connectivity between the fracture planes, had to be removed and masked carefully before moving on to the next step (**Fig**. **7**). Several techniques such as the application of cluster detection algorithms and geometric interpolation were utilized in this process. Neighboring voxels with similar properties were grouped together by cluster detection. First, all of the voxels within a selected range or properties were labeled. One of the labeled voxels was selected, recorded and given a new label. Then all of its nearest neighboring voxels of the newly labeled voxel were recorded and given a new label. In the next step, all of the nearest neighbors of the relabeled sites (with the exception of previously relabeled voxels) were relabeled and recorded, and the process was repeated until no more voxels with the origin label could be found. The total list of recorded sites is a cluster. A new voxel with the original label was found, and the entire process was repeated to find the second cluster. The entire procedure was repeated as often as required until all of the clusters were located, and all of the initially labeled voxels were relabeled (Tout et al., 2007).

### 3.2.3 Image binarization and segmentation of the cracks

A number of methods have been developed for segmentation of "foreground" objects from a "background" through simple thresholding, edge detection, active contour shape detection and indicator kriging. The work presented here begins once the images have been converted to binary form to isolate various objects, i.e. the cracks, the pores, and the background solid matrix. These methods are described in-depth in the scientific literature (Lindquist et al., 2000). A simple thresholding procedure was used to convert the 8-bit grey-scale images into binary form. The pore spaces and volumes were segmented using a specific range of grey-values between 0 and 35 (out of 256 grey levels), where zero represents black (which corresponds to pores, which have no X-ray attenuation) (**Fig. 8a**). The value of 35 was selected visually and by trial and error with fracture plane continuity as the goal. Besides, it was not possible to distinguish real features from artifacts if a threshold below 35 was selected. Choosing the threshold value of 35 introduces some errors, however selecting values less than 35 results in artificially disconnected pore spaces, which do not adequately represent fracture planes. A threshold value of 35 includes voxels that represent volumes with microcracks smaller than 5 µm in diameter. When these smaller microcracks are excluded (i.e., use smaller threshold values), the fracture planes are disjointed. By including these smaller microcracks (< 5 µm), reconstruction of the fracture plane with greater continuity is possible. Thresholding was followed by noise-reduction filtering to make sure that visible cracks and other features do not disappear after this step (**Fig. 8b**). This was necessary and helped save a considerable amount of computation time in the next step.

In the next phase of this study we will be interested in studying transport phenomena inside the material and transport through its boundaries. Therefore, the connectivity of the pore spaces and crack apertures and their connectivity with the exterior must be captured and preserved. To preserve the fracture planes a region-based segmentation method called the watershed method was utilized to form fracture planes from seemingly isolated, but densely grouped voxels with grey-values approaching zero **(Figs. 8c** through **8g)**. This method was originally proposed by Lantuéjoul (1978). Contrary to simple cluster identification, separations obtained using the watershed algorithm are much more representative of the local geometry. The watershed method finds the "ridge lines" between neighboring "basins" (Vincent and Soille, 1991). One minor inconvenience, also reported by others (Cendre et al., 1999), is that the separation is just one-voxel thick, while our goal is to assign every voxel of the pore space to a pore. This necessitates post-processing to assign these watershed voxels to an adjacent pore volume and smooth the fracture planes. The final product, despite all the filtering, is still a more-or-less continuous medium with small amplitude heterogeneity filled with a number of small width crack apertures. To make sure that "cracks" artificially generated during image acquisition and processing were not interpreted as real cracks, only cracks larger than a certain size (10000 connected voxels, below which the crack density drops by 3 orders of magnitude) were selected for quantitative description in terms of crack thickness, roughness, orientation, spatial correlation, displacement and the strain field. Obviously this filters out some of the real, but small fractures, but a statistical study of the crack size distribution was not part of this study (**Fig 9**). In this figure, filtering of some large scale and oriented cracks from many small and scattered ones is shown (at the current resolution it is difficult to determine any patterns formed by the small cracks.). The large scale fractures are clearly orientated along the plane of the lamination, which is perpendicular to the coring direction.

### 4. Discussion
Digital correlation analysis (DIC) of both optical (Rechenmacher and Finno, 2004; Bornert et al., 2010) and X-ray (Louis et al., 2007; Lenoir et al., 2007; Viggiani, 2009) images is now quite widely used in experimental mechanics, particularly in geo-materials mechanics, and it is applied to 2D (Rechenmacher and Finno, 2004; Hild and Roux 2006; Louis et al., 2007; Bornert et al., 2010) and 3D (Lenoir et al., 2007 ; Viggiani 2009) images. The fact that the 2D digital image correlation analysis of Green River Shale during heating was successful indicates that uncontrolled rigid body motion of the sample was sufficiently small that correlation between one image and the next was not lost. Similarly, the rigid body motion was small enough for the strain fields (the derivatives of the displacement field) to be determined.

Digital image correlation allowed us to quantify the evolution of the linear thermal expansion coefficient with temperature. Depending on the direction relative to the bedding plane, the linear thermal expansion coefficient is typically 2 to 5 times smaller in the horizontal direction, parallel to the bedding plane, than in the vertical direction. Depending on the temperature, we found vertical linear thermal expansion coefficients between about $2\times10^{-5}$ and $20\times10^{-5}$ /$^{O}$C. These values may be compared with, and they were found to be in good agreement with, the reported *volumetric* thermal expansion coefficients of the kerogen in Green River shales and for the mineral matrix in which it is embedded. The latter is reported to be in the range $2\text{-}5\times10^{-5}$/$^{O}$C (see e.g. (Settari 2001; Chen, 2003)), whereas the former is in the range $10\times10^{-5}\text{-}30\times10^{-5}$/$^{O}$C below 145$^{O}$C (the glass temperature of the kerogen) and in the range $55\times10^{-5}\text{-}60\times10^{-5}$/$^{O}$C above 145 $^{O}$C (Zhang, 2009).

The spatial resolution of the measured displacement fields is limited by the size of the sub-images used for correlation (of the order of ten pixels, i.e. 50µm). At each pixel, the measured displacement has a typical resolution of a tenth of the pixel size, i.e. 500nm. The maximum displacement distance that can be accurately measured with the correlation analysis approach used here is around one quarter of the sub-image size (6 pixels for sub-images of size 25 pixels) so that larger displacements were discarded (the proportion of discarded pixels was less than 1.6% at all temperatures below 350 $^{O}$C). As mentioned above, this prevented us from correlating most of the images with the initial image to determine total displacements directly. In principle, larger displacements could be measured directly, by changing the correlation procedure, but this would be at the expense of a higher computation cost.

In this work, we limited the correlation analysis to 2D slices of the volume image and did not attempt a computationally demanding 3D correlation analysis. In practice, this was not a limitation for the study of the macroscopic strain before the onset of fracturing since the behaviors determined for the two orthogonal slices extracted from the 3D volume were almost identical, and they are therefore representative of the behavior of the whole volume. After fracturing began, open cracks significantly altered the images thus preventing application of the correlation analysis described in this paper, be it 2D or 3D. In this case, 3D image analysis proved to be an efficient tool to monitor crack propagation and coalescence. In future work, 3D correlation analysis could, in principle, be used to determine the locations of crack nucleation sites and identify precursors to crack nucleation once the cracks or their precursors become large enough to result in a voxel gray-scale value equal to or less than 35. In reality undetected nanocracks (and even smaller scale damage) are likely to be microcrack precursors. However, we anticipate that it would be difficult to distinguish small-scale/small-amplitude precursors from the imaging/correlation noise. As a consequence, we used 2D correlation analysis before crack nucleation and we then employed 3D image analysis to investigate crack nucleation and growth.

X-ray CT can be used for quantitative and qualitative analysis of internal features only if those features are revealed by sufficiently great differences in atomic composition and/or density. Pronounced variations in X-ray attenuation coefficients are required to distinguish two substances in the CT images. A strong contrast exists between solid phases and air filled pore volumes, and in our case there was a clear contrast between the cracks and mineral phases. However, the contrasts between other components are smaller, and our experimental/synchrotron setup did not allow us to distinguish between kerogen-containing pore spaces (kerogen lenses) and cracks filled with gas/kerogen mixtures.

Utilizing the 4D CT image analysis results, it can be seen that individual microcracks propagate via small local advances in the crack front, which are controlled by the heterogeneity and evolving stress/strain field. The nature of the crack propagation and the geometries of both individual microcracks and the ensemble of microcracks generated in the heating experiment suggest that the process that generates them might be related to percolation or invasion percolation. The standard percolation model is physically realistic since kerogen is distributed more or less randomly along bedding planes throughout the organic rich rock and decomposes to produce fluids everywhere. However, it would be very surprising if the evolving stress field played no role in the process. A modified percolation or invasion model might be more realistic (**Fig. 10**). The nucleation and growth of cracks in heterogeneous media due to the internal production of fluids is a complex phenomenon that has not been investigated in detail. Phase separation of gas dissolved in a liquid can be brought about by changing the composition (the gas/liquid ratio), the pressure or the temperature. In simple liquids, more or less spherical gas filled bubbles are formed via nucleation and growth during the early stages of phase separation (while the gas/liquid volume ratio is small). However, if the phase separation takes place in a gel, the gas filled volumes are more like penny shaped cracks than spherical bubbles (Renard et al., 2009; Boudreau et al., 2005), and under these conditions, bubbles and cracks are indistinguishable. In particular, the production of gases in mud that consist of a three-dimensional gel network of fine particles, which penetrate a volume of aqueous fluid, results in the formation of penny-shaped gas-filled cracks that can be seen in X-ray tomography images (Boudreau et al., 2005). In the case of kerogen rich rocks, hydrocarbon fluids are formed inside the kerogen particles, and fracturing could be initiated by phase separation inside the kerogen. The molecular architecture of the type I kerogen in the Green River Shale can be thought of as a three-dimensional gel-like (cross-linked polymer) network. However, bubble/crack nucleation could occur on kerogen/mineral interfaces or the interfaces between liquid filled pores and minerals or their interfaces between liquids and kerogen. The X-ray images suggest the growth of bubbles or cracks in a heterogeneous medium. It is surprising that the cracks are not more like penny-shaped cracks when they are small, but the Green River shale is much more heterogeneous than synthetic gels or mud, particularly on scales greater than the gel network correlation length (typically smaller than 1 micrometer).

Although the possibility that small cracks might be present in the materials that we investigated or could have been created when the specimens were collected or when cores and thin sections were prepared cannot be ruled out, but no evidence was found for such fractures in either the thin sections or the cores used in the present study.

In the heating experiment, the sample was heated continuously for 340 minutes as the temperature rose from 60ºC to 400 ºC at the constant rate of 1 ºC /minute. It took 20 minutes to image each sample, and 17 images were recorded during the heating stage. An additional 51 images (images 18 - 68) were acquired every 20 minutes at a constant temperature of 400 ºC. Out of the inspected images quite large number of very small cracks appeared in image 23, and by image 28 out of the 68 image sequence, these microcracks had grown and coalesced into much larger cracks (**Fig. 10**). The small sizes of the cracks particularly when they first appear, increases the challenge of robust segmentation. To distinguish between the kerogen patches, which were originally present, and the evolving fluid-filled cracks, it is important to use information from the preceding images. Without this information, it is not possible to differentiate between kerogen and other forms of hydrocarbons generated via kerogen decomposition because they have similar X-ray attenuation coefficients. Even higher tomographic resolutions would be required to investigate whether cracks nucleate within kerogen patches, at kerogen-mineral interfaces or elsewhere. Again, comparison of successive images would be needed to distinguish between cracks and pores, which have characteristic dimensions smaller than the size of the voxels used in the work reported here. Image correlation analysis could also be used to indirectly detect early stage cracks with sub-voxel scale apertures, via the displacements associated with them.

A possible explanation for the peak in the strain curve at ~120°C (**Fig. 4**) is dehydration of clay minerals such as smectites. The $2^{nd}$ peak can be explained by the volume change due to kerogen decomposition, and other processes such as oxidation of organic matter may occur when the temperature reaches ≈320°C.

The crack aperture decrease when the crack front reaches the boundary of the sample (**Fig. 10c**). This suggests that the pressure generated by the low molecular weight kerogen decomposition products inflates the cracks, and that this pressure is released when the cracks reach the exterior. This observation also supports the hypothesis that fracturing is caused by fluid generation, and that the excess pressure inside the cracks drives crack growth. The cracks all have more-or-less the same orientations, which are apparently controlled by the strongly anisotropic sedimentary morphology (i.e. lamination). The TEM, XRF and SEM images indicate that some elements are concentrated on the laminations. Damage was observed to initiate at multiple locations within the same interface between two laminations, and the microfractures (damage zones) grow within an interface. Then damaged regions expand and finally interconnect to form a fracture that spans the entire sample. Obviously under subsurface conditions, except at very small depths, the confinement might keep the fracture closed or the aperture might not be large enough to be visible at micron scale resolution. Another issue that has not been considered here is the correspondence of hydrocarbon escape rate to the time-resolution of the X-ray tomography experiment.

One of the main objectives of these experiments was to investigate the applicability of time-lapse high resolution synchrotron CT to study the evolution of a model source rock during heating. The next step will be to study the same process under confinement – under conditions more similar to those that prevail during natural primary migration. Of course, the time scale of any practical laboratory experiment will be shorter than the time scale of natural kerogen decomposition by many orders of magnitude. This will make fracturing more likely in the experiments because there will be much less time for the fluids produced by kerogen decomposition to migrate via flow or diffusion. However, we do expect that experiments under confinement will lead to important insights into the mechanisms of natural primary migration. The time scales associated with the commercial production of hydrocarbon fluids from oil shales, such as the Green River Shale, are much shorter (hours for surface retorting and years for in situ, subsurface, retorting). The results reported here are directly relevant to surface retorting, and experiments with small confining stresses (0.2 – 20 MPa) would be relevant to in situ production.

## 5. Conclusion

4D (three space dimensions + time) X-ray computed tomography was performed using synchrotron beam lines which made it possible to acquire images at resolutions of a few microns with short scanning time among other benefits (monochromatic beam to avoid beam hardening, etc.) We have shown how the deformation and fracturing of shales during heating can be investigated using X-ray tomography, and a workflow was developed to extract very small crack apertures from microtomography images of tight-rocks samples. This workflow could be modified and customized to extract other features for other geomaterials and other processes that lead to deformation or fracturing and deformation. A conceptual model was presented, which explains why and where these cracks nucleated, how they grew and finally merged together. It has been shown how three-dimensional X-ray tomography detection and analysis of fractures can be supported by two-dimensional correlation analyses.

For the Green River Shale sample that was used to investigate the growth of cracks under heating and without confinement, the following observations can be made:

- On micrometer scales, cracks nucleation, growth and coalescence does not occur randomly. Instead, it was observed that cracks nucleate and grow in regions with the highest material heterogeneities in which the highest stress concentrations also occur. Expansion of the organic material present in the shale samples and its transformation into gas explains the stress build up. Individual cracks grow in the bedding plane in which they are located until they meet other crack or reach the boundary of the shale sample. The aperture widths of the cracks do not change substantially when they grow, but once the cracks reach the sample boundaries their aperture widths decrease.

- On larger scales, a set of parallel fractures grow, and the common orientation of the fractures is controlled by the natural lamination.
- The strain increases homogeneously along both the *x* and *y* directions. However the strain is found to be anisotropic, and the horizontal strain increment (along the stratification) is significantly smaller (by a factor of about 2 to 5 depending on the temperature) than the vertical strain (perpendicular to the stratification).
- The approach taken here, which combines 2D image analyses with 3D image analyses, has the advantage of compensating for the shortcomings of each approach. The quantitative 2D analysis was complemented by qualitative 3D analysis and it was shown that the 2D results are characteristic of the 3D behavior.

**Acknowledgements**: We would like to thank Anne-Céline Ganzhorn, Berit Osteby, Elodie Boller at ESRF, and Samuel McDonald at SLS for their help during the acquisition of tomography data. Rodica Chiriac from LMI laboratory at the University Claude Bernard in Lyon performed the thermogravimetry analyses.

# Tables

| Analyzed Sample(sample name) | TC-TOC-TIC/Geochemical Analyses | | |
|---|---|---|---|
| | TC | TOC | TIC |
| Green River Shale-01 | 13 wt% | 10 wt% | 3 wt% |
| Green River Shale-02 (GRSII, GRSIV) | 13 wt% | 9 wt% | 4 wt% |

Table 1- Two geochemically-analyzed Green River Shale Samples for which X-ray microtomographic images were constructed. These samples were acquired from the Green River Formation, USA. TC: Total Carbon, TOC: Total Organic Carbon, TIC: Total Inorganic Carbon.

**Figures**

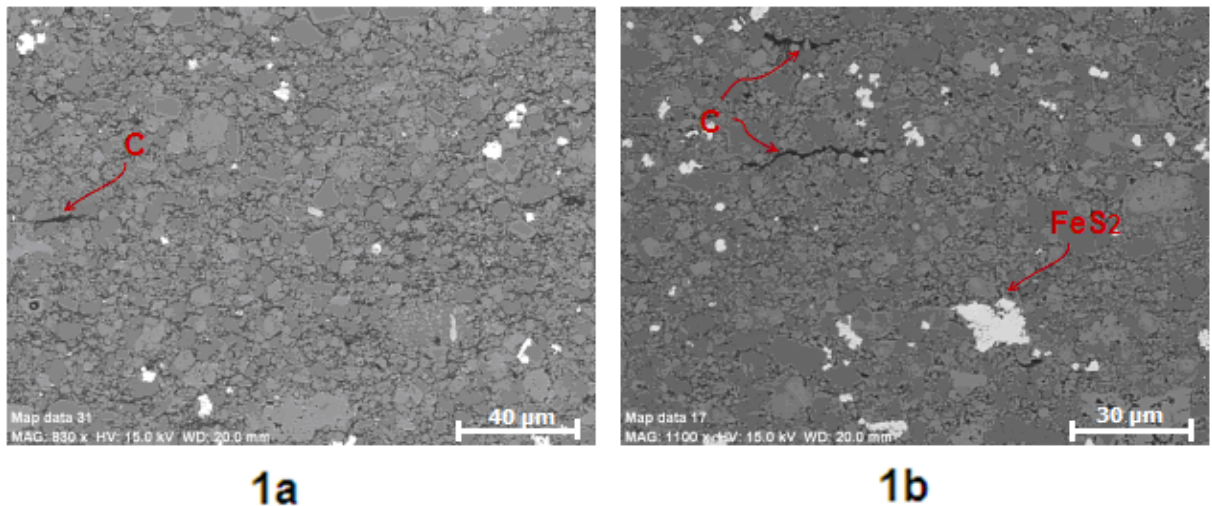

Fig. 1- Parts a and b display back-scattered electron images of two Green River Shale thin-sections. The atomic backscattering intensity increases with increasing atomic number, and the greatest backscattered electron intensity (the most lightly shaded areas) correspond to dense minerals such as pyrite composed of high atomic number minerals. The darkest areas correspond to low density, low atomic number, components such as organic materials, and the black regions indicate voids.

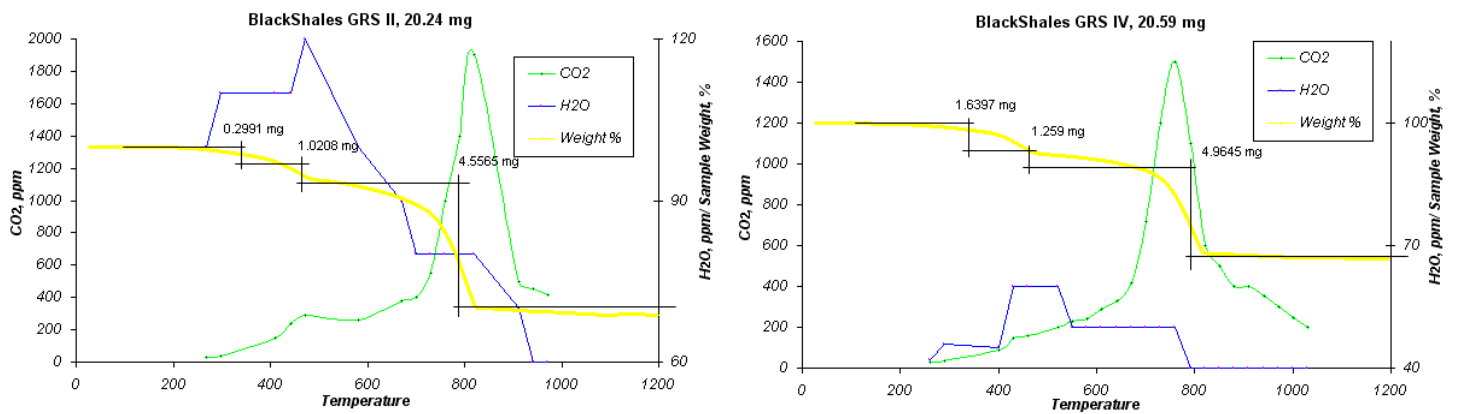

Fig. 2- Thermogravimetric analyses of Green River Shale samples. Each figure displays the weight loss (yellow curve), carbon dioxide release (green curve) and water release (blue curve). Three main steps of mass release could be identified. From left to right step 1 corresponds to dehydration of clay minerals. Step 2 corresponds to the temperatures over which hydrocarbons are released and detected by gas chromatography. Step 3 corresponds to decomposition of carbonate minerals at temperatures between 750°C and 850°C.

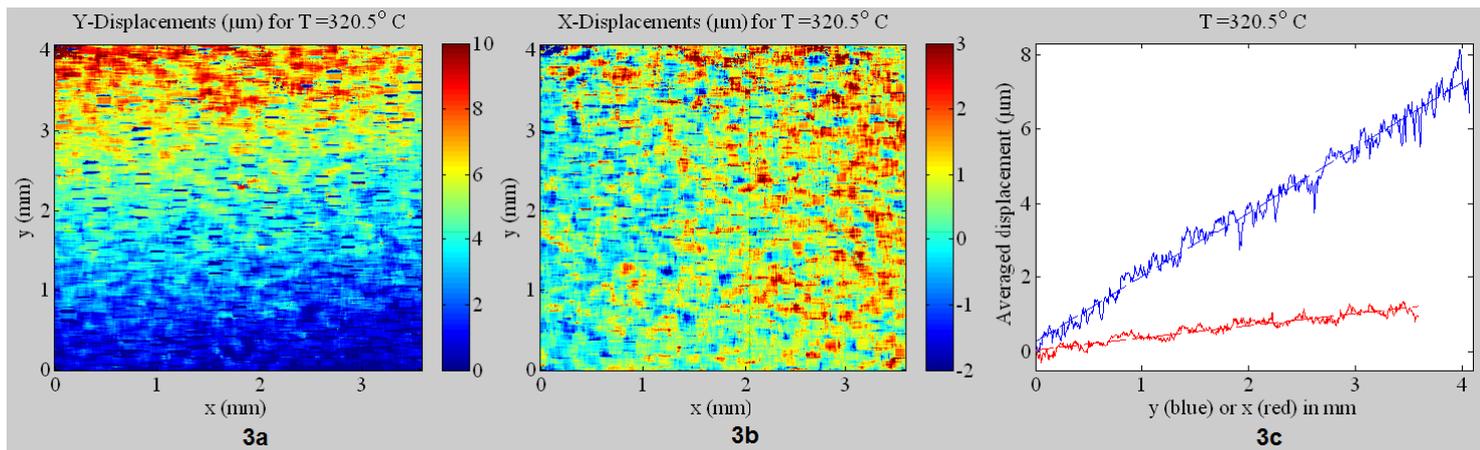

Fig. 3- Displacement fields measured by digital image correlation (DIC) between two successive images around T=320.5°C, using a sub-image area of 25 X 25 pixels. 3a- color map of the spatial distribution of the vertical (y direction) displacements in micrometers (the bottom most line has an average vertical displacement of 0). 3b- color map of the spatial distribution of the horizontal (x direction) displacements in micrometers (the leftmost line has an average horizontal displacement of 0). 3c- Displacement averaged along x (resp. y) for the y (resp. x) displacement as a function of y (resp. x).

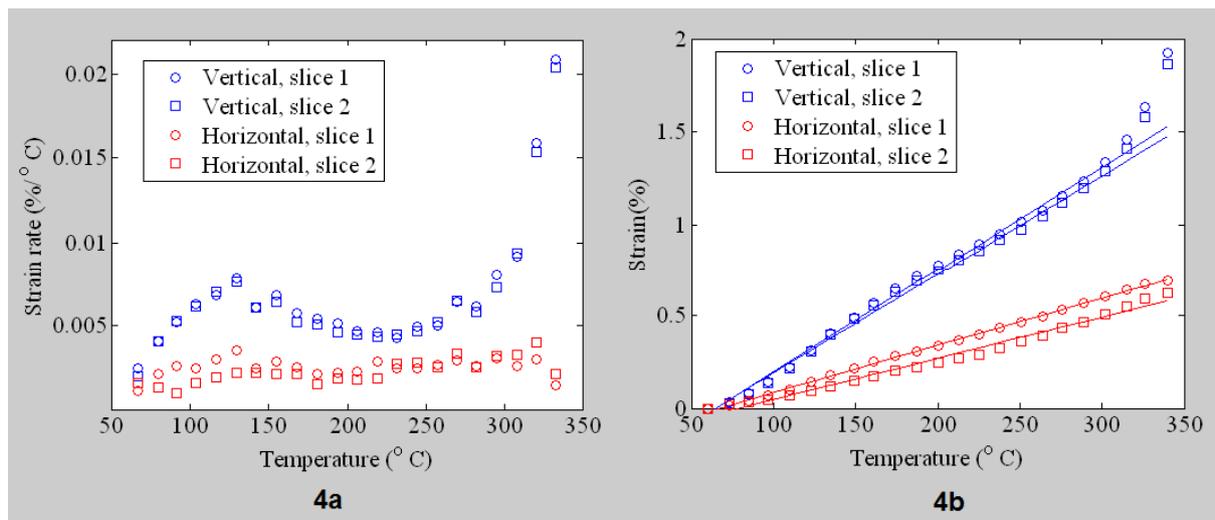

Fig. 4- 4a: Evolution of the strain increment between two images. 4b: Strain as a function of temperature, for both directions (vertical in blue, horizontal in red) and for both slices. The vertical strain − temperature curve deviates from linearity at temperatures of about 320°C, which corresponds to the onset of fracturating.

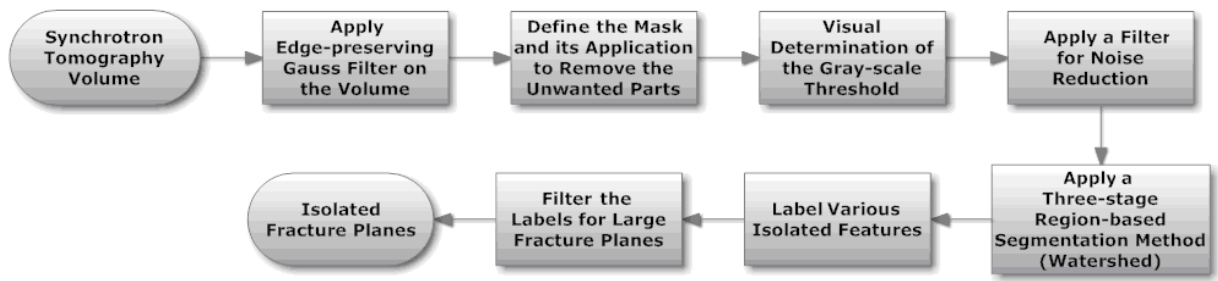

Fig. 5- Processing workflow diagram for the 3D tomogram.

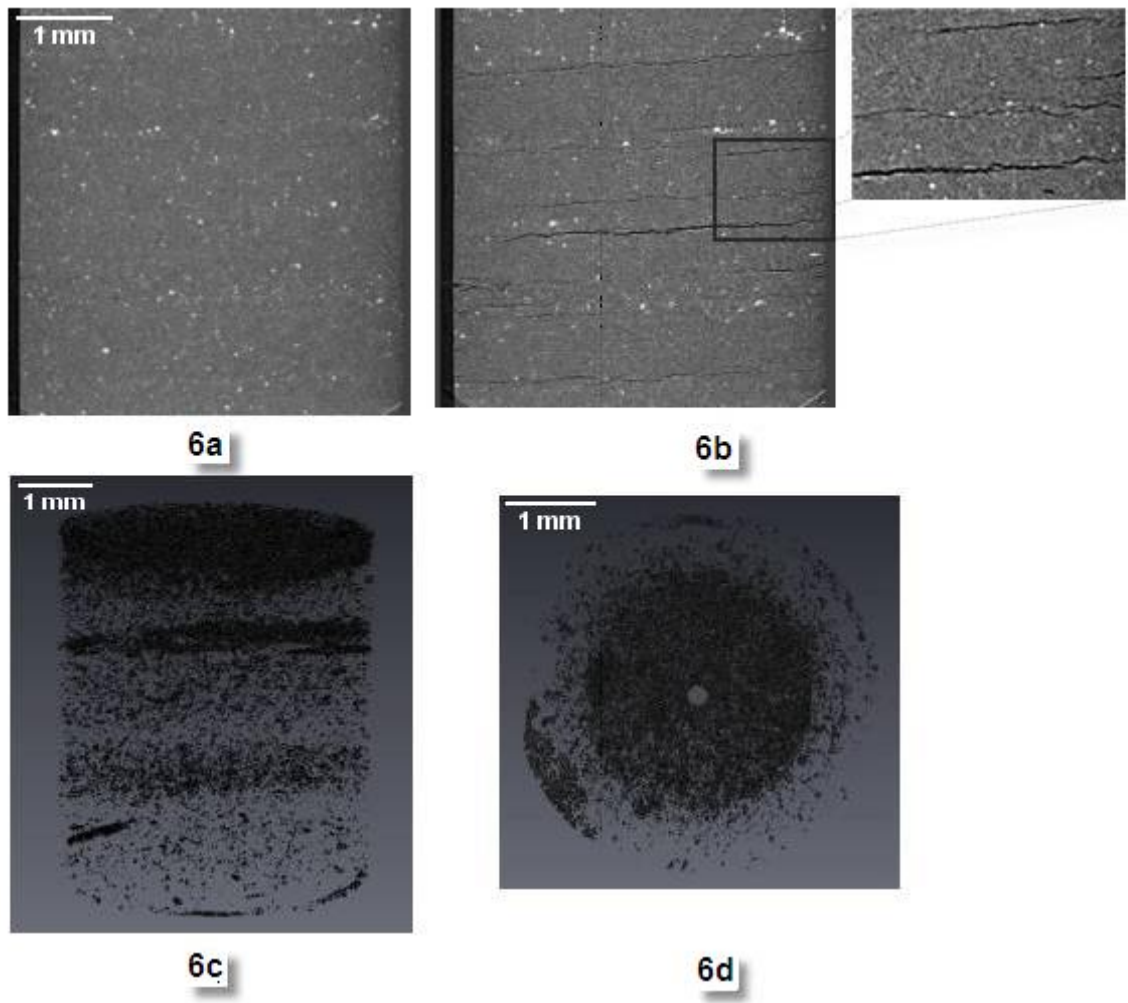

Fig. 6- Microtomography images of Green River Shale. 6a- 2D slice at the onset, and 6b- after heating (4 mm x 4 mm) showing that cracks began to appear at about 320°C. 6c and 6d- 3D views of porosity showing small crack apertures (in the range 1-5 voxels/ 5-25 micrometer) and a wide range of crack sizes, which makes the segmentation procedure challenging.

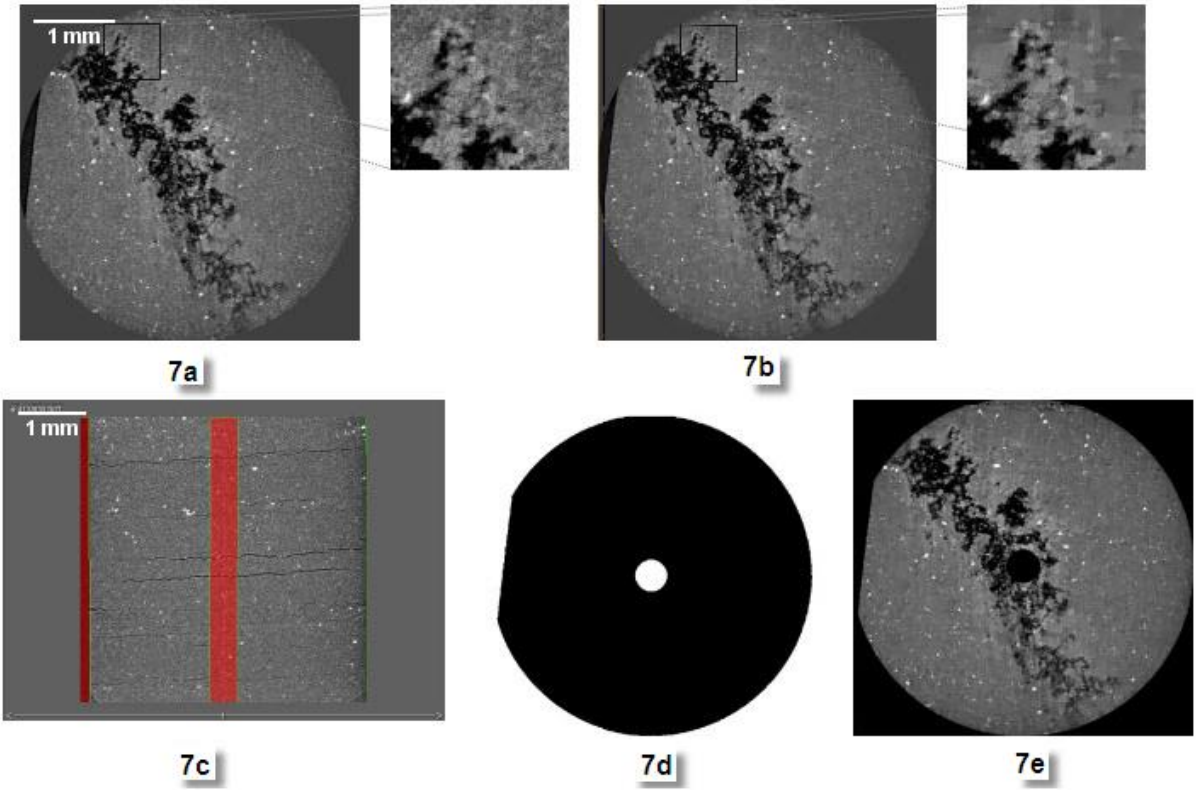

Fig. 7- Primary filtering and masking workflow. 7a- Original volume showing a crack (dark area). 7b- Gaussian filtered volume. 7c- Definition of the mask (red areas) representing the areas which should be taken out of the processing. 7d- The mask, and 7e- Masked and filtered volume. The centre of the sample and an edge, where the noise was higher than in the rest of the sample, were removed.

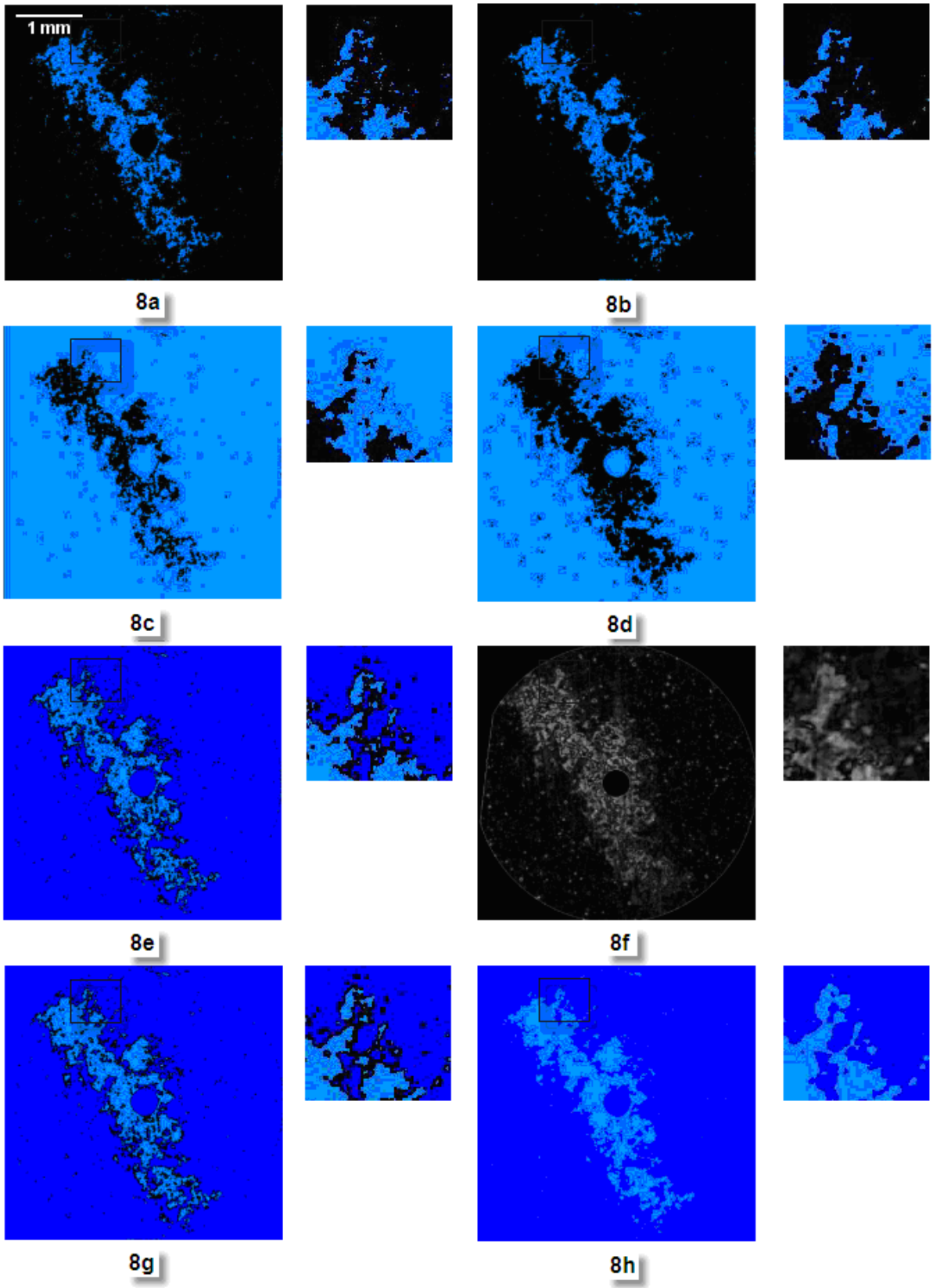

Fig. 8- Image processing workflow. 8a- Thresholded volume. 8b- Noise-reduced volume. 8c- Pre-watershed markers volume 0. 8d- Pre-watershed markers volume 1. 8e- Watershed markers volume. 8f- Gradient volume of the original volume. 8g- Watershed volume. 8h- Dilated watershed volume.

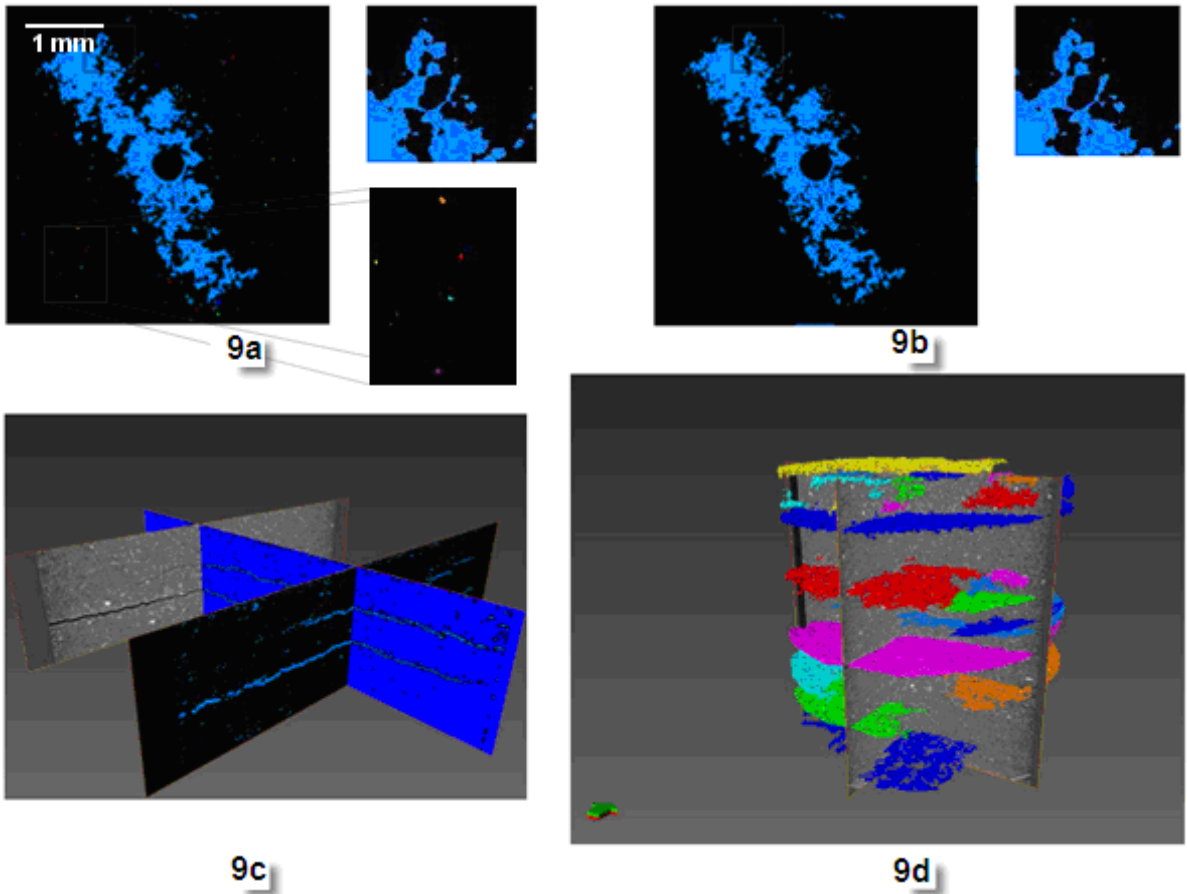

Fig. 9- Post processing workflow. 9a- Labeled volume. 9b- Filtered labeled volume. 9c- Isolated digitized cracks. 9d- Filtered and labeled fracture planes.

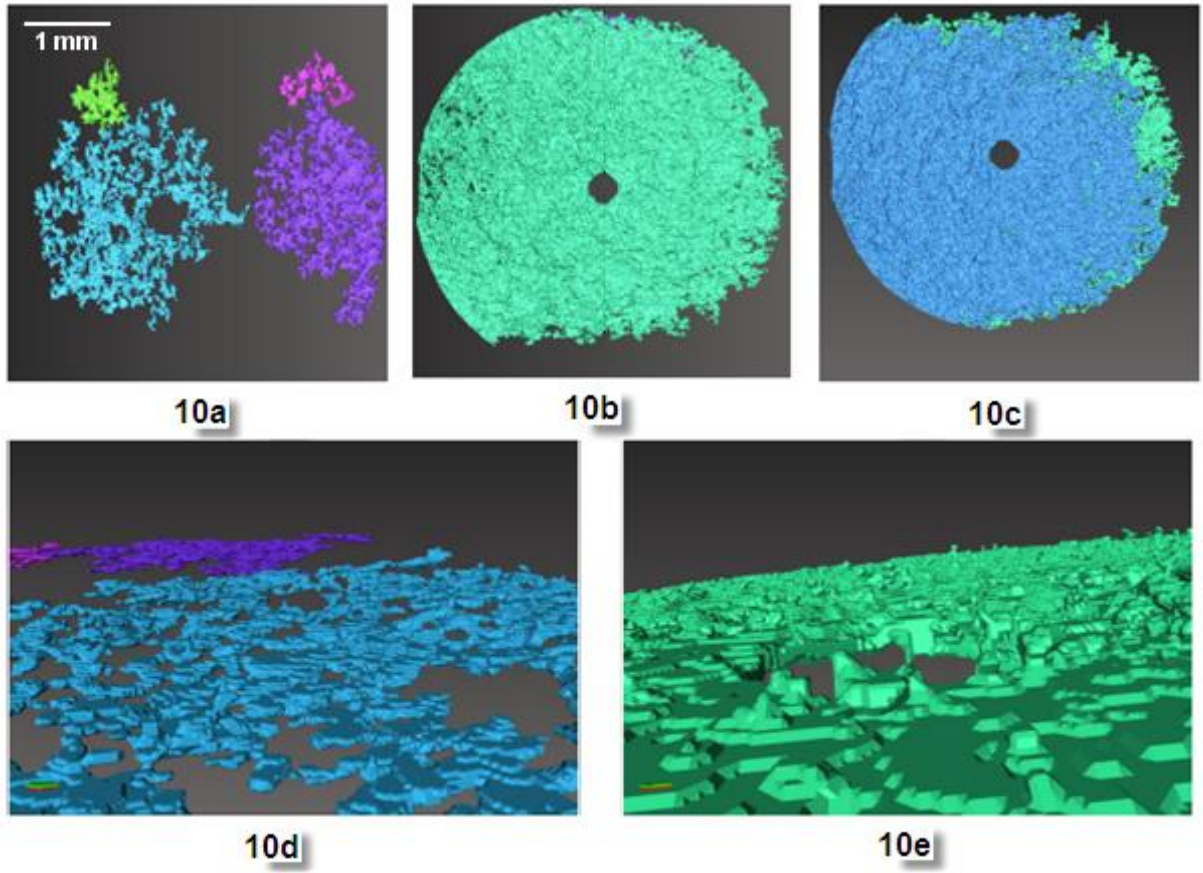

Fig. 10- Imaging of a single fracture. 10a- Crack appearance at an early stage of growth; 10b- Cracks merging into one another; 10c- Overprint of the crack long after reaching the boundary in blue on the one at the onset of hitting the boundary in green; 10d- Cracks show out-of-plane fluctuations. 10e- Surface roughness of the merged cracks.